\documentclass[aps,prl,twocolumn,superscriptaddress,
nobibnotes,nodoi,noeprint,longbibliography]{revtex4-2}
\usepackage{graphicx}
\usepackage{dcolumn}
\usepackage{float}
\usepackage{color}
\usepackage{lmodern}
\usepackage{bm}
\usepackage{amsmath,amssymb}
\usepackage{mleftright} 
\usepackage{dsfont}
\usepackage{mathtools} 
\usepackage{siunitx}
\usepackage{graphicx}
\usepackage{amssymb}
\usepackage{graphicx}
\usepackage{ bbold }
\usepackage{pgf}
\usepackage{gensymb}

\renewcommand\matrix[1]{\,\bm{\underline{#1}}\,}
\renewcommand\epsilon{\varepsilon}
\renewcommand\rho{\varrho}
\renewcommand\vec[1]{{\boldsymbol #1}}

\renewcommand\geq\geqslant
\renewcommand\leq\leqslant

\usepackage{mathrsfs}
\usepackage{esint}
\usepackage{ulem}
\usepackage[unicode=true,pdfusetitle,bookmarks=true,
bookmarksnumbered=false,bookmarksopen=false,breaklinks=false,
pdfborder={0 0 0},backref=false,colorlinks=true,citecolor=red]
{hyperref}
\begin{document}
\title{Photocatalytic magnetic microgyroscopes
with activity-tunable precessional dynamics}
\author{Dolachai Boniface}
\affiliation{Departament de F\'{i}sica de la Mat\`{e}ria Condensada, Universitat de Barcelona, 08028 Spain}
\author{Arthur V. Straube}
\affiliation{Zuse Institute Berlin, Takustra{\ss}e 7, 14195 Berlin, Germany}
\affiliation{Freie Universit{\"a}t Berlin, Department of Mathematics and Computer Science, Arnimallee 6, 14195 Berlin, Germany}
\author{Pietro Tierno}
\email{ptierno@ub.edu}
\affiliation{Departament de F\'{i}sica de la Mat\`{e}ria Condensada, Universitat de Barcelona, 08028 Spain}
\affiliation{Universitat de Barcelona Institute of Complex Systems (UBICS), Universitat de Barcelona, Barcelona, Spain}
\affiliation{Institut de Nanoci\`{e}ncia i Nanotecnologia, Universitat de Barcelona, Barcelona, Spain}
\date{\today}
\begin{abstract}
Magnetic nano/microrotors 
are passive elements that spin around an axis due to an external rotating field while remaining confined to a close plane. They have been used to date in different applications 
related to fluid mixing,  drug delivery or biomedicine. 
Here we realize an active version of 
a magnetic microgyroscope which is 
simultaneously driven by 
a photo-activated catalytic reaction and a rotating magnetic field.
We investigate the uplift dynamics of this colloidal spinner when it stands up and precesses around its long axis while self-propelling due to the light induced decomposition of hydrogen peroxide in water.
By combining experiments with theory,  we show that activity 
emerging from the cooperative action of phoretic and osmotic forces  effectively increase the gravitational torque which counteracts the magnetic and viscous ones,
and carefully measure its contribution.
\end{abstract}
\maketitle
Spinning tops are mechanical
anisotropic objects able of defying gravity while spinning around their vertical axis on a tiny tip~\cite{Perry1957}.
When actuated by a twisting force, the tops 
can raise at a vertical position and precess slowly around an axis 
until friction and dissipation inevitably terminate their motion.
Spinning tops are also part of mechanical gyroscopes, 
important  for navigational systems~\cite{Britting1971,Passaro2017} and used as
precise inertial sensors~\cite{Perry1957}.
Thus, the macroscopic precession 
of anisotropic systems has been a subject of extensive research to date.

Recent years have witnessed an increasing interest in investigating 
the spinning motion of driven micro/nanoscale  
particles, due to their 
direct application in disparate technological fields
including microfluidics~\cite{Sawetzki2008,Kavcic2009,Chong2013}, microrheology~\cite{Berret2016,Gu2016}, sensor~\cite{Steimel2014,Kiran2023} and  
biotechnology~\cite{Xianglong2023}.
Experimental realizations 
of precessing magnetic stirrers at such scale 
include
ferromagnetic nanorods~\cite{Dhar2007} or
magnetic Janus colloids~\cite{Yan2012,Gao2015},
and have been recently used to investigate the  entropy-driven 
thermal re-orientation of a single element~\cite{Gao2015} or
the collective self-assembly process~\cite{Mecke2023}. However, all these cases involve passive particles in absence of self-propulsion, which prevent  the rotating element from moving and stirring across the plane.
%
\begin{figure}[!ht]
\begin{center}
\includegraphics[width=0.95\columnwidth,keepaspectratio]{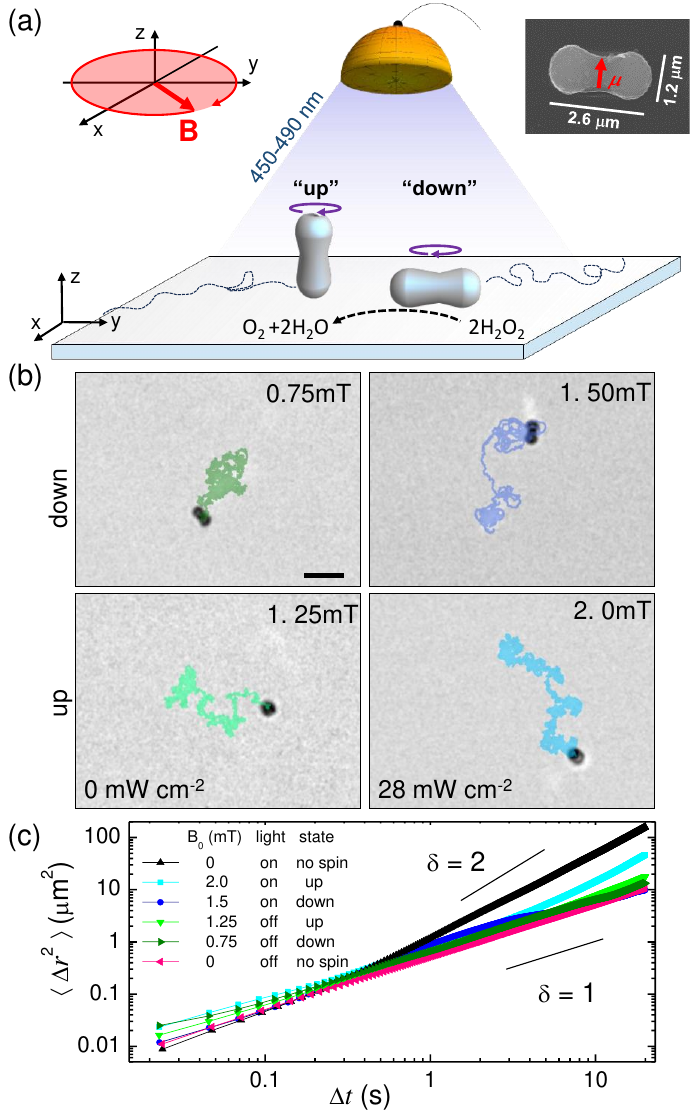}
\caption{(a) Schematic showing the two dynamic states, ``up'' and ``down'', of a hematite particle under the combined action of a rotating magnetic field $\bm{B}$ 
and the light induced decomposition of hydrogen peroxide (H$_2$O$_2$) in water. 
Small inset at the top right shows scanning electron microscope image of one hematite particle 
with the permanent moment $\bm{\mu}$ superimposed.
(b) Optical microscope images of 
the hematite particle in the down (top row) and up (bottom row) states in absence (first column) and in presence (second column) 
of blue light with the trajectories overlayed. The number at the top of each image shows the
field amplitude of the rotating field, scale bar in the first image is $5 \,\rm{\mu m}$. The corresponding video (Video S1) can be found in the Supporting Information. 
(c) Translational mean squared displacement $\langle \Delta r^2 \rangle $
versus lag-time $\Delta t$ for spinning rotors in different dynamic states.}
\label{figure1}
\end{center}
\end{figure}

In this context, ferromagnetic nanorods have been used in the past
as microrheological tools~\cite{Anguelouch2006,Petesic2011,Tokarev2012,Chevry20132,Brasovs2015,Radiom2021}, 
or to create localized microvortices able to trap~\cite{Mair2011,Petit2012,Torres2017} and stir~\cite{Keshoju2007,Chevry2013} non-magnetic tracer particles. 
Introducing activity via self-propulsion may lead to rich dynamic states with additional functionality. For example, an active magnetic rotor will be able to translate, and does not need to stop its rotational motion within the dispersing medium.
As a matter of fact, passive micro/nano stirrers lack 
of translational motion, which could only be induced by changing the plane of rotation of the applied field, i.e. when it 
rotates perpendicular to the sample plane~\cite{Petit2012} such that these particles stop stirring and behave as surface rotors~\cite{Zhang2010,Tierno2021,Junot2023}. 

Here we realize an active magnetic microgyroscope that combines two key elements: on one hand, it functions as a passive microgyroscope that spins due to a rotating magnetic field, and on the other, it operates as an active system exhibiting self-propulsion induced by an independent photoactivated chemical reaction. 
By illuminating with blue light, we measure an increase in the threshold field required to uplift and demonstrate that activity 
which results from phoretic and osmotic forces, enhances the gravitational torque effect by also pushing the particles toward the close plane.
These results align well
with a theoretical model that captures the main physical mechanisms governing the uplift
dynamics.

Our active microgyroscopes are made of 
ferromagnetic hematite particles with a peanut shape, as shown in the small inset in Figure~\ref{figure1}(a). This particular shape results from the
sol-gel process used to synthesize them~\cite{Lee2009}. The  fabricated particles are 
characterized by two connected lobes with a long (short) axis
equal to $a = 2.6 \,\rm{\mu m}$ ($b = 1.2 \,\rm{\mu m}$).
As observed in a previous work~\cite{Palacci2013},
hematite colloids become active when subjected to blue light in a water solution of 
hydrogen peroxide (H$_2$O$_2$).
Their surface exposed to blue
light catalyzes the decomposition of H$_2$O$_2$ in water,  following the reaction:
$2$H$_2$O$_{2(l)}\rightarrow$ O$_{2(g)}$+2H$_2$O$_{(l)}$. 
This reaction creates a chemical concentration gradient around the particles,
inducing a diffusiophoretic flow. Due to this flow, the hematite particle can attract passive colloids nearby~\cite{Pedrero2017} and display self-propulsion~\cite{Palacci2013b}.

We disperse the particles in an aqueous  basic solution containing $3.6\%$ by vol. of H$_2$O$_2$, which is enclosed within a rectangular glass micro-tube (inner dimension 2$\times0.1\,$mm). After a few minutes, the hematite 
particles sediment nearby to the bottom plate due to density mismatch, and display there diffusive dynamics. 
We induce 
self-propulsion by applying blue light at a wavelength $\lambda = 450-490\,$nm with a tunable intensity $I\in [0,125] \, \rm{mW \, cm^{-2}}$. Moreover, before the experiments, we employ an etching treatment with hydrochloric acid  to enhance the particle activity~\cite{Palacci2013b}. 
Additionally, we apply an external rotating magnetic
field to induce spinning motion. 
This is possible since the hematite 
particles are slightly ferromagnetic, 
and display a small permanent dipole moment $\mu =2 \times 10^{-16} \,\rm{A\, m^2}$
oriented  along their 
short axis~\cite{Pedrero2016},  Figure~\ref{figure1}(a).
More technical details on the experimental protocol and setup can be found in Section 1 of the Supporting Information.

We apply a rotating magnetic field 
circularly polarized in a plane $(\hat{\bm{x}}, \hat{\bm{y}})$ parallel to the glass substrate:
\begin{equation}
\bm{B} = B_0[\cos{(\Omega t )} \hat{\bm{x}} - \sin{(\Omega t )} \hat{\bm{y}}] \, ,
\label{field_equation}
\end{equation}
with $B_0$ the field amplitude, $\Omega=2 \pi f$, where $f$ is the driving frequency.
The hematite particles perform a spinning motion around their 
short axis with a frequency $f_s$, and for sufficiently low frequencies 
they rotate parallel to the bounding plane, which we refer to as the ``down'' state.
In contrast, at intermediate frequencies,
the particles preferentially stand up, preforming a spinning motion around their 
long axis, thus perpendicular to the glass substrate, the ``up'' state, see Fig.~\ref{figure1}(a).
These two dynamic states can be controlled by adjusting $B_0$ and $f$, 
in addition to that we include the light activation, making these spinning rotors self-propelling.  
Fig.~\ref{figure1}(b) displays typical trajectories of the hematite rotors in four possible situations: when they are in the down (top row) and up (bottom row)  states and in presence of light (right column) or in absence of it (left row). 

In absence of light, $I=0$, the rotating hematite particles perform standard diffusive dynamics standing up or laying down, as shown by the measured translational mean squared displacement (MSD) in Fig.~\ref{figure1}(c).
Here we calculate the MSD as, $\langle \Delta \bm{r}^2 (\Delta t) \rangle \equiv \langle (\bm{r}(t)-\bm{r}(t+\Delta t))^2 \rangle \sim \Delta t^{\delta}$, with $\bm{r}(t)$ the position of the particle center 
at time $t$,
$\Delta t$ the lag time and $\langle \dots \rangle$ a time average. The MSD can be used to distinguish between the normal diffusive ($\delta = 1$) dynamics from the sub-[super] diffusive ($\delta < 1$ [$\delta > 1$]) and ballistic ($\delta = 2$) ones.
When the particles are activated by light ($I=28 \, \rm{mW \, cm^{-2}}$) we find that 
in absence of spinning ($B_0=0$), 
the active particle exhibits a ballistic trajectory 
after few seconds with $\delta \sim 2$. In contrast, the spinning effect localizes the trajectories, reducing the corresponding exponent in the MSD. Specifically, we observe diffusive behavior for spinning down and super-diffusive, almost ballistic behavior for spinning up.
This effect can be understood by observing that a rotating object in a viscous fluid 
generates an hydrodynamic flow field~\cite{Happel1983}, which in first approximation is purely azimuthal and decays
as $\sim 1/r^2$~\cite{Lenz2003}.
Such flow may transport away the product 
of the catalytic reaction, reducing thus  
the concentration gradient around the particle 
and thereby the corresponding self-propulsion~\cite{Das2015,Uspal2016,Ketzetzi2020}. 

%
\begin{figure}[!ht]
\begin{center}
\includegraphics[width=0.95\columnwidth,keepaspectratio]{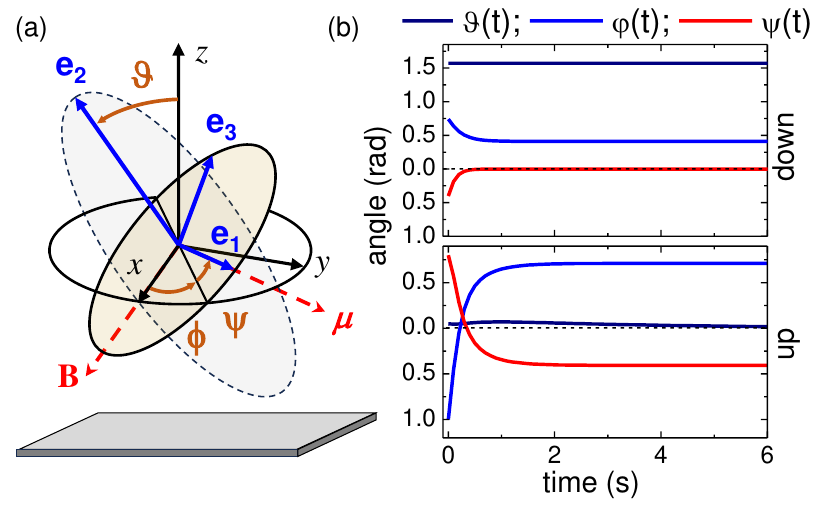}
\caption{(a) Schematic showing an elliptical particle (dashed line) 
with the laboratory reference frame ($\hat{\bm{x}}$,$\hat{\bm{y}}$,$\hat{\bm{z}}$), the coordinate system 
fixed on the particle ($\hat{\bm{e}}_1$,$\hat{\bm{e}}_2$,$\hat{\bm{e}}_3$), the three Euler angles ($\vartheta,\phi,\psi$), and the orientation of the applied magnetic field $\bm{B}$ and magnetic moment $\bm{\mu}$. (b) Time evolution of the Euler angles from Eq.~\eqref{eq:main} for $\hat{B}=5.0$
and at two different driving frequencies, $\hat{\Omega} = 2.0$ (down state, top panel, note $\vartheta=\pi/2$) 
and $\hat{\Omega} = 3.0$ (up state, bottom panel, note $\vartheta=0$).}
\label{figure2}
\end{center}
\end{figure}

To gain insights in the different dynamic states, we have modified the model suggested in Ref.~\cite{Dhar2007} to our situation, by incorporating the additional effect of the catalytic reaction and the different geometry (shape). We approximate the particle as an ellipsoid with the major axis $a$ and minor axis $b$,
mass $m$, permanent moment $\bm{\mu}$ oriented along the minor axis and suspended in liquid medium of dynamic viscosity $\eta$.
Further, as shown in Fig.~\ref{figure2}(a), we 
have used two coordinate systems, the laboratory one ($\hat{\bm{x}}$,$\hat{\bm{y}}$,$\hat{\bm{z}}$) 
and a second one that is fixed to the 
ellipsoid ($\hat{\bm{e}}_1$,$\hat{\bm{e}}_2$,$\hat{\bm{e}}_3$) with $\hat{\bm{e}}_1$ and $\hat{\bm{e}}_3$ aligned with $\bm{\mu}$ and the long axis of the ellipsoid, respectively. Both reference frames are connected through the  Euler angles, ($\vartheta,\phi,\psi$). 
In the absence of chemical activity, the overdamped motion of the ellipsoid is governed by the balance of different torques: magnetic, $\bm{\tau}_m=\bm{\mu} \times \bm{B}$,  gravitational, $\bm{\tau}_g=ma(\hat{\bm{e}}_3 \times \bm{g})/2$, and viscous, $\bm{\tau}_{\eta}=-\matrix{\zeta} {\bm{\omega}}$, torques:
\begin{align}
    \matrix{\zeta} {\bm{\omega}} = \bm{\mu} \times \bm{B}+ \frac{1}{2} m a (\vec{e}_3 \times \vec{g})\,, \label{eq:torque}
\end{align}
where $\bm{\bm{g}}$ is the gravitational acceleration, 
 $\bm{\omega}$ is the angular velocity and $\matrix{\zeta}=\text{diag}(\zeta_{1},\zeta_{2},\zeta_{3})$, $\zeta_{1}=\zeta_{2}\neq\zeta_{3}$ is the rotational friction tensor in the main axes.
Our main hypothesis is that the chemical activity generates an attractive phoretic/osmotic force $\bm F_{p/o}$ between the hematite and the substrate. This force gives rise to a corresponding torque,
$ \bm\tau_{p/o}=a(\hat{\bm{e}}_3 \times \bm{F}_{p/o})/2$
which prevents the hematite from uplifting inducing an effective increase of the gravitational torque torque $\bm\tau_\downarrow=\bm\tau_{p/o}+ \bm\tau_g$. Also, since the the hematite is pressed against the substrate there is a raise in the drag torque which now becomes $\bm \tau_d$. To take into account the effect of these osmotic/phoretic contributions avoiding the 
complex details of the corresponding chemical 
reactions, we introduce the dimensionless prefactor $\alpha \geq 1$, $\beta \geq 1$, 
\begin{align*}
 \bm\tau_\downarrow \to \alpha \bm\tau_g \,, \quad \bm\tau_d \to \beta \bm\tau_\eta \,
\end{align*}
such that $\alpha=1$, $\beta=1$, for the system without activity.

Introducing dimensionless time, $\hat{t} = {mga}/({2\beta \zeta_{1}}) t$, frequency, $\hat \Omega = {2\zeta_{1} \Omega}/({mga})$, and field, $\hat B = {2 \mu B_0}/({mga})$, 
we arrive at the governing equations of motion:
\begin{subequations} \label{eq:main}
    \begin{align}
        \dot \vartheta & = \left(\alpha + \hat B \sin\varphi \sin\psi\right) \sin\vartheta \,, \label{eq:main-theta} \\ 
        \dot \varphi & = \beta \hat \Omega - \hat B \sin \varphi \cos \psi\,, \label{eq:main-varphi} \\
        \dot \psi & = - \frac{\hat B}{1-\kappa}\left(
        \kappa\sin\varphi\cos\theta\cos\psi + \cos\varphi \sin\psi\right)\,. \label{eq:main-psi}
    \end{align}
\end{subequations}
Here, $\kappa=(\zeta_{1}-\zeta_{3})/\zeta_{1}$ specifies the particle's asymmetry and we have introduced the angle $\varphi=\phi-\Omega t$, implying the transition to the reference frame rotating together with the external field. The cases with $\vartheta=\pi/2$ and $\vartheta=0$ describe the down and up states, respectively. For $\alpha=1$, $\beta=1$, Eq.~\eqref{eq:main} corresponds to the passive system \cite{Dhar2007}, when the blue light and hence the activity are switched off. Figure~\ref{figure2}(b) shows that Eq.~\eqref{eq:main} correctly predicts the two dynamic states
experimentally observed by varying the field parameters. For a fixed value of the rescaled field, $\hat{B}=5.0$ the stable state is the 
 rod laying down ($\vartheta=\pi/2$) for low frequency, $\hat{\Omega}=2.0$ (top panel), or standing up ($\vartheta=0$) for large one, 
$\hat{\Omega}=3.0$ (bottom panel).

\begin{figure}[!ht]
\begin{center}
\includegraphics[width=\columnwidth,keepaspectratio]{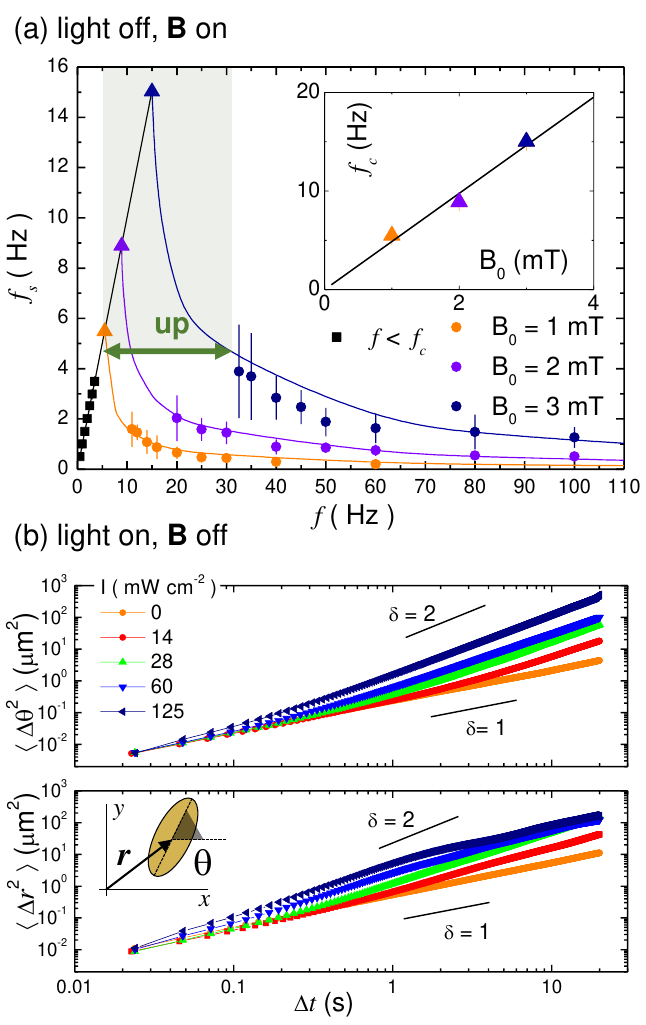}
\caption{(a) Spinning frequency $f_s$ of a single hematite particle versus 
field frequency $f$ for three field amplitudes $B_0$. Scattered symbols are experimental
data, continuous lines are non-linear regressions of the synchronous ($f<f_c$)
and asynchronous ($f > f_c$ ) regimes. The triangles indicate the values of $f_c$ 
extracted from these curves, and are shown in the top inset. 
The shaded green region in the graph indicates that the particle is in the up state. (b,c) Angular $\langle \Delta \theta^2 \rangle$ (b)
and translation $\langle \Delta r^2 \rangle$ (c) mean squared displacements 
of a hematite particle in absence of magnetic field and under different 
light amplitudes $I$.}
\label{figure3}
\end{center}
\end{figure}

We first characterize with experiments  how the
particle dynamics are influenced independently   by 
the magnetic field (Fig.~\ref{figure3}(a)) and by the activity (Fig.~\ref{figure3}(b,c)). At low frequencies we observe the down state. Such state, where $\vartheta=\pi/2$, $\psi=0$ is described by the single simplified equation, 
$\dot \varphi  = \beta \hat \Omega - \hat B \sin \varphi$, cf. Eq.~\eqref{eq:main-varphi}.  The latter equation predicts two dynamic regimes separated by a critical frequency $f_c$. Formulated in the dimensional units relative to the laboratory reference frame, for $f<f_c$ the particle rotates synchronously with $\bm{B}$,
and its average spinning frequency $\langle f_s  \rangle=  f$,
as shown by the black squares in Fig.~\ref{figure3}(a). In contrast, for $f>f_c$ the particle still spins but does it in an asynchronous regime, where 
the average spinning drops down as, 
$\langle f_s\rangle= f [1-\sqrt{1-(f_c/f)^2}]$,
see the disks in Fig.~\ref{figure3}(a). 
These two regimes are connected by the critical frequency 
$f_c$ which, as shown in the small inset in Fig.~\ref{figure3}(a),
scales linearly with the applied field as $f_c=\mu B_0/(2\pi\zeta_1)$.
The data in such inset have been extrapolated from the 
non-linear regressions, since for a range of intermediate frequencies  the spinning hematites were observed to enter in the up state rather than remained confined to the close plane. This effect is highlighted by the shaded green region in the graph in Fig.~\ref{figure3}(a).

Our magnetic rotors can also be  activated by blue light. In Fig.~\ref{figure3}(b), we show the effect of the light induced decomposition of H$_2$O$_2$, by measuring both the 
angular MSD $\langle \Delta \theta^2 \rangle$ (top) and the translational one, $\langle \Delta \bm{r}^2 \rangle$ (bottom).
In absence of light ($I=0$), both MSDs show diffusive dynamics with a long-time rotational diffusion coefficient $D_{\theta}=0.140 \pm 0.001 \, \rm{rad^2 \, s^{-1}}$ and a translational one, $D_{\bm{r}}=0.152 \pm 0.001 \, \rm{\mu m^2 \, s^{-1}}$.  However, when the light is on, the hematite particle becomes activated and displays a super-diffusive dynamics.
This effect strongly emerges in the angular dynamics, 
where we initially observe a 
diffusive behavior followed by a 
superdiffusive/ballistic one at large $\Delta t$.
The threshold between both dynamic states decreases with the light amplitude $I$,
and for the maximum power ($I=125 \, \rm{mW \,cm^{-2}}$) it reduces to $\Delta t \sim 0.1\,$s.  
The effect of activity also changes the translational MSDs. Below a light intensity of $I\sim 50 \, \rm{mW \, cm^{-2}}$, we observed a diffusive behavior comparable to the case without activity, which then becomes superdiffusive beyond $\Delta t=1\,$s. In contrast, for $I> 50 \, \rm{mW \, cm^{-2}}$ the behavior is first superdiffusive, then returns to diffusive for $\Delta t >3-5\,$s. In this case we observe an enhanced diffusion with a coefficient which is more than $10$ times higher than the passive case,  $D_{\bm{r}}\sim 2 \, \rm{\mu m^2 \, s^{-1}}$.
This feature results from the etching process 
of the particles (see Method section), which roughens the particle surface enhancing the generated chemophoretic flow and the corresponding self-propulsion behavior. 

\begin{figure}[!t]
\begin{center}
\includegraphics[width=\columnwidth,keepaspectratio]{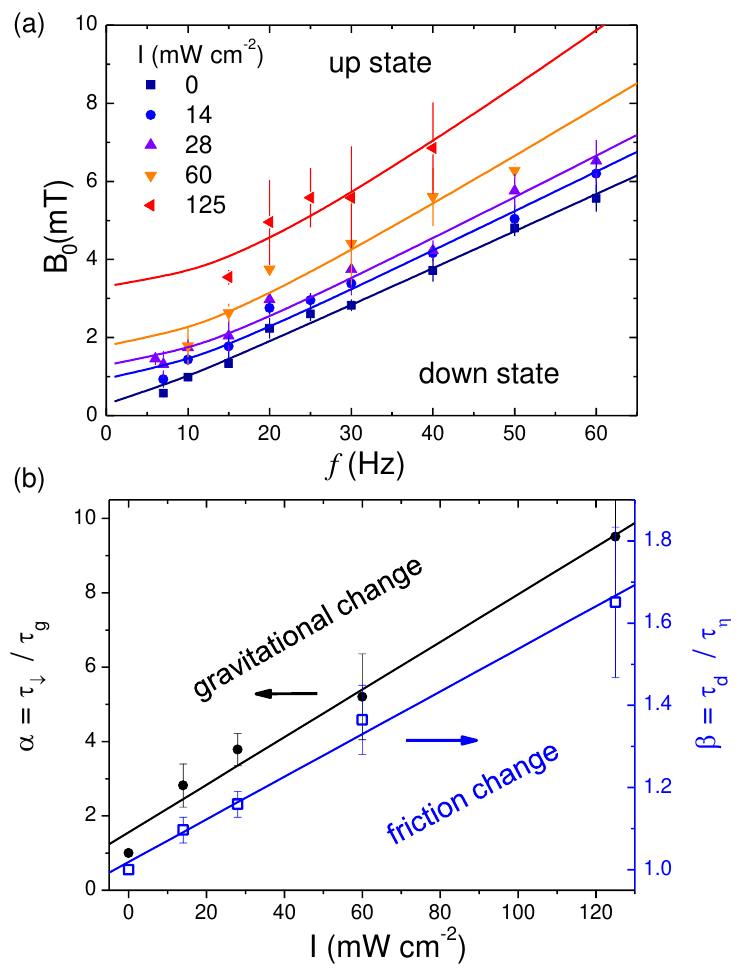}
\caption{(a) Border between the up and down states in the $(f,B_0)$ plane and for different light intensities $I$. 
Scattered symbols are experimental data while continuous lines are non-linear regression from Eq.~\eqref{threshold}.
(b) Evolution of the parameter $\alpha$ (black) indicating the raise of the gravitational-like torque ($\bm{\tau}_g$) due to the light intensity $I$, 
and $\beta$ (blue) which corresponds to the variation of the friction torque ($\bm{\tau}_{\eta}$). }
\label{figure4}
\end{center}
\end{figure}

We now consider the combined effect of the 
activity and spinning on the particle dynamics. 
The catalytic reaction induces self-propulsion due to a local gradient triggering osmotic and phoretic phenomena. Also, this reaction creates
an effective attractive force between the hematite and the substrate.
This force tends to elevate the threshold field $\hat{B}_*$ required for the particle to transit towards the up state. We confirm this effect
in Fig.~\ref{figure4}(a), where we have measured  $\hat{B}_*$ 
by varying the light intensity, and thus the activity, 
in the $(B_0,f)$ plane. Thus, the blue light shifts the limit $\hat{B}_*$ towards higher magnetic amplitudes. 
From the model given by Eq.~\eqref{eq:main}, it follows that 
the up and down state start to coexist at a critical field value, 
\begin{equation}
    \hat B_*(\hat\Omega) = \frac{1}{\alpha}\sqrt{(\alpha^2 + \beta^2 \hat \Omega^2)(\alpha^2+\kappa^2\beta^2 \hat \Omega^2)} \,.
\label{threshold}
\end{equation}
The complete derivation of this equation is given in Section 2 of the Supplementary Material.
Eq.~\ref{threshold} shows that, at low $\kappa$, changing $\alpha$ will  shift the critical magnetic field amplitude, while changing $\beta$ will alter the slope of the curve for sufficiently small $\Omega$. In the latter case, we specifically verify this effect by increasing the viscosity of the solution through the addition of glycerol. In this specific scenario, we confirm that $\beta = \eta / \eta_{\text{w}}$, where $\eta$ is the viscosity of the solution and $\eta_{\text{w}}$ that of water, as shown in Section 3 of the Supplementary  Information.

We use Eq.~\eqref{threshold} to fit the experimental data 
describing the uplift transition induced by 
the rotational motion, Fig.~\ref{figure4}(a). 
First we consider the transition in absence of activity, $I=0$. Thus, 
we fix $\alpha=1$, $\beta=1$ and determine the remaining experimental parameter, here $\kappa=0.0076$ and the rescaling prefactors. 
Next, we fix $\kappa$ 
and change both $\alpha$ and $\beta$ performing the simultaneous regression of all the experimental data obtained at different $I$.
The multiple regression confirms the good agreement between the model and the experimental data.
More importantly, this agreement highlights that the effect of activity effectively increases the the vertical force pushing the particle towards the substrate. Even if the attraction is towards the nearby substrate, the presence of an uplift force arising from the 
magnetic torque allows us to precisely measure this contribution.
Indeed,
from the variation of $\hat B_*(\hat\Omega)$ it follows that $\alpha$, which represent the intensity of attractive
force relative to gravity, increases linearly with the light intensity. This phoretic/osmotic $\bm F_{p/o}$ force 
scales  the gravity from up to $10$ times for the maximum light intensity, $I=125 \, \rm{mW \, cm^{-2}}$ and the
light induced torque $\bm{\tau_\downarrow} \propto \bm{\tau}_g$,
Fig.~\ref{figure4}(b).
Concerning the quantity $\beta$, which can be considered as a measure of the variation of the drag
torque, it also increases with the light intensity
but less drastically, as shown by the blue line in Fig.~\ref{figure4}(b). This effect is due to the fact that the attractive phoretic force pushes the hematite close to the substrate, thus raising the hydrodynamic drag coefficients. 

We finally comment on the origin of the chemically induced 
force $\bm F_{p/o}=\bm F_p + \bm F_o $, which results 
from both diffusiophoresis ($\bm{F}_p$) and diffusioosmosis ($\bm{F}_o$) contributions. 
The first one arises from a surface concentration gradient along the hematite surface, while the second is due to a concentration gradient along the substrate. For both contributions, the surface concentration gradient $\nabla_\parallel \phi$, along the hematite or substrate, generates a slip velocity $u_\text{slip} \propto \bm{\nabla_\parallel} \phi$, which leads to forces acting on the hematite~\cite{Boniface2024} due to the generated hydrodynamic flow fields. Both components tend to push  the particle 
toward the surface. The phoretic effect has been observed in previous works~\cite{Palacci2013,Massana2018}  to induce an attraction between passive objects, as the substrate, and  photocatalytic active hematite particles. For the osmotic component, the chemical activity of the hematite generates a radial concentration gradient along the substrate. The slip velocity over the substrate initiates a centripetal osmotic flow~\cite{Boniface2024}. Due to the incompressibility of water, a vertical flow compensates for the radial flow, pushing the hematite against the substrate.

In conclusion, we have experimentally realized an active magnetic microgyroscope 
where spinning and activity are simultaneously and independently controlled by two different 
actuating fields. 
We investigate how spinning affects the microgyroscope transport and find that the rotational motion strongly suppresses 
self-propulsion compared to the zero magnetic field case.
We then investigate the uplift dynamics when these gyroscopes
are subjected to an in-plane rotating field and 
stand up to reduce viscous dissipation.
In particular, 
by balancing all torques acting on the 
rotating particles, we find that the activity 
induced by a catalytic chemical reaction can be considered 
as an additional, gravitational-like torque, which increases the amplitude threshold field 
to transit the particle in the up state. 
We use a theoretical model to capture the basic physics of the process by balancing magnetism, gravity, viscosity and activity. 
These spinning self-propelling agents can be used in microfluidic channels as active component,
adding further feasibility to previous experimental
realizations based on passive (i.e. non-active) colloidal rotors~\cite{Terray2002,Tierno2008}.

\begin{acknowledgments}
We thank Arkady Pikovsky for helpful discussions.
This project has received funding from the 
European Research Council (ERC) under the European Union's Horizon 2020 research and innovation programme (grant agreement no. 811234).
A. V. S. acknowledges support by the Deutsche Forschungsgemeinschaft (DFG) under Germany’s Excellence Strategy–MATH+: The Berlin Mathematics Research Center (EXC-2046/1)–Project No. 390685689. 
P. T. acknowledges support from the 
Ministerio de Ciencia, Innovaci\'on y Universidades (grant no. PID2022-
137713NB-C21 AEI/FEDER-EU) and the Ag\`encia de Gesti\'o d'Ajuts Universitaris
i de Recerca (project 2021 SGR 00450) and the Generalitat de Catalunya (ICREA Acad\'emia).
\end{acknowledgments}

\end{document}